\documentclass[JOURNAL,10pt]{IEEEtran}

\usepackage{color}
\usepackage{float}
\usepackage{tabularx,colortbl}
\usepackage{setspace}
\usepackage{cite,graphicx,amsmath,psfrag,bm}
\usepackage{subfigure}
\usepackage[usenames,dvipsnames]{xcolor}
\usepackage{mathabx}
\usepackage{cancel}
\usepackage{epstopdf}
\usepackage{amsmath}
\usepackage{extarrows}
\usepackage{enumitem}

\usepackage[process=auto]{pstool}
\usepackage[normalem]{ulem}
\graphicspath{{Figs/}}

\newcommand{\figref}{Fig.~\ref}

\ifCLASSINFOpdf
\else
\fi

\begin{document}

\title{Microwave Hilbert Transformer and its Applications in Real-time Analog Processing (RAP)}

\author{Xiaoyi~Wang,~\IEEEmembership{Student~Member,~IEEE,}
        Zo\'{e}-Lise~Deck-L\'{e}ger,~\IEEEmembership{Student~Member,~IEEE,}
        Lianfeng~Zou,
        Jos\'{e}~Aza\~{n}a,
        and Christophe Caloz,~\IEEEmembership{Fellow,~IEEE}
        % <-this % stops a space
\thanks{ Xiaoyi Wang, Lianfeng Zou, Zo\'{e}-Lise Deck-L\'{e}ger and Christophe Caloz are with department of electrical engineering, Polytechnique Montr\'{e}al, Montr\'{e}al, Qu\'{e}bec, H3T 1J4, Canada. xiaoyi.wang@polymtl.ca}
\thanks{Jos\'{e}~Aza\~{n}a is with Institut National de la Recherche Scientifique -- \'{E}nergie Mat\'{e}riaux et T\'{e}l\'{e}communications (INRS-EMT),
Varennes, Qu\'{e}bec, J3X 1S2, Canada.}
}

\maketitle

\begin{abstract}
A microwave Hilbert transformer is introduced as a new component for Real-time Analog Processing (RAP). In contrast to its optical counterpart, that resort to optical fiber gratings, this Hilbert transformer is based on the combination of a branch-line coupler and a loop resonator. The transfer function of the transformer is derived using signal flow graphs, and two figures of merits are introduced to effectively characterize the device: the rotated phase and the transition bandwidth. Moreover, a detailed physical explanation of its physical operation is given, using both a steady-state regime perspective and a transient regime perspective. The microwave RAP Hilbert transformer is demonstrated experimentally, and demonstrated in three applications: edge detection, peak suppression and single sideband modulation.
\end{abstract}


\section{Introduction}
\emph{Real-time Analog Processing (RAP)} is a microwave-terahertz-optical technology that consists in manipulating electromagnetic signals in real-time using dispersion-engineered components called \emph{phasers}~\cite{Jour:2013_MwMag_Caloz}. This technology has recently been introduced as a potential high-speed and low-latency alternative to dominantly digital technologies, given its unique features of real-time operation, low-power consumption and low-cost production~\cite{Jour:2013_MwMag_Caloz}. RAP phasers have been realized in different architectures, including C-sections and D-sections~\cite{JOUR:2010_TMTT_Gupta,JOUR:2014_JRMCAE_Zhang,JOUR:2015_TMTT_Gupta,JOUR:2012_MWCL_Horii,Conf:Wang_URSIGASS2017}, coupled resonators~\cite{JOUR:2012_TMTT_Zhang,JOUR:2013_TMTT_QZhang}, nonuniform delay lines~\cite{JOUR:2003_TMTT_Laso_FourierTransform,JOUR:2003_TMTT_Laso}, metamaterial transmission-line structures~\cite{JOUR:2009_TMTT_Abielmona}, loss-gain pairs~\cite{Jour:Zou_2016_MTT_Loss-Gain}, and RAP has been demonstrated in several applications, including compressive receiving~\cite{JOUR:2009_TMTT_Abielmona}, real-time spectrum analysis~\cite{JOUR:2009_TMTT_Gupta,Jour:Wang_2017_RLSD}, real-time spectrum sniffing~\cite{JOUR:2011_TMTT_Nikfal,Jour:Wang_2018_sniffer}, Hilbert transformation~\cite{Conf:Wang_APS2018_HT}, SNR enhanced impulse radio transceiving~\cite{JOUR:2014_MWCL_Nikfal}, dispersion code multiple access (DCMA)~\cite{Jour:Zou_TWC_2018_DCMA}, signal encryption~\cite{Conf:Wang_APS2017_Phaser}, radio-frequency identiﬁcation~\cite{JOUR:2011_AWPL_Gupta} and scanning-rate control in antenna arrays~\cite{JOUR:2017_TAP_Zhang_PhaserScanningArray}.

Efficiently performing \emph{mathematical operations} is central to any type of processing. Whereas mathematical operations are completely natural in digital signal processing, they are much less obvious in RAP, where they are still essential to help making this technology more flexible and efficient. Fortunately, several real-time analog mathematical operations have already been successfully demonstrated, including differentiation~\cite{Jour:Caloz_2008_MWCL_Differentiator}, integration~\cite{Jour:Yao_2016_NP_Reconf_interg_SP}, expansion and compression~\cite{JOUR:2009_TMTT_Abielmona}, time reversal~\cite{CONF:2008_IRWS_Schwartz}, Fourier transformation~\cite{JOUR:2003_TMTT_Laso_FourierTransform} and Hilbert transformation~\cite{Jour:Azana_2014_OL_HT_Filter}.

The Hilbert transformation is a particularly fundamental operation in both physics and engineering. It has already been implemented in a few RAP applications, such as single side-band (SSB) modulation~\cite{Jour:Yao_2016_NP_Reconf_interg_SP,Jour:Tanaka_2002_EL_SSB_HT}, band-pass and band-stop filtering~\cite{Jour:Azana_2014_OL_HT_Filter} and edge/peak detection~\cite{Jour:Davis_2000_OL_edge_detect}. However, the Hilbert transformers reported to date have been exclusively designed in the optical regime using optical fiber structures that are not transposable to lower frequencies, and they  have been mostly described in purely mathematical terms with little insight into the device physics. This paper reports a \emph{RAP microwave Hilbert transformer}, based on branch-line couplers and resonant loops, and explains the physical operation of this transformer, using both time-domain and steady-state perspectives.

\section{Recall on the Hilbert Transformation}
\begin{figure}[h!]
    \centering
    \includegraphics[width=0.98\columnwidth]{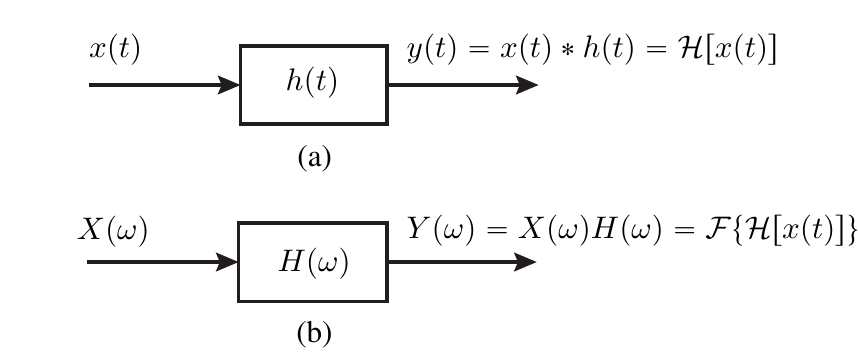}
        %\psfragfig[width=0.98\columnwidth, trim={0in 0in -1.2in 0in}]{Figs/HTSche}{
%        \psfrag{A}[c][c][0.9]{$x(t)$}
%        \psfrag{B}[c][c][0.9]{$X(\omega)$}
%        \psfrag{C}[c][c][0.9]{$h(t)$}
%        \psfrag{D}[c][c][0.9]{$H(\omega)$}
%        \psfrag{E}[l][l][0.9]{$y(t)=x(t)*h(t)=\mathcal{H}[x(t)]$}
%        \psfrag{F}[l][l][0.9]{$Y(\omega)=X(\omega)H(\omega)=\mathcal{F}\{\mathcal{H}[x(t)]\}$}
%        \psfrag{a}[c][c][0.9]{(a)}
%        \psfrag{b}[c][c][0.9]{(b)}
%        }
        \caption{Hilbert transformation. (a)~Time domain. (b)~Frequency domain.}
   \label{FIG:HTSche}
\end{figure}

\begin{figure}[h!]
    \centering
    \includegraphics[width=0.98\columnwidth]{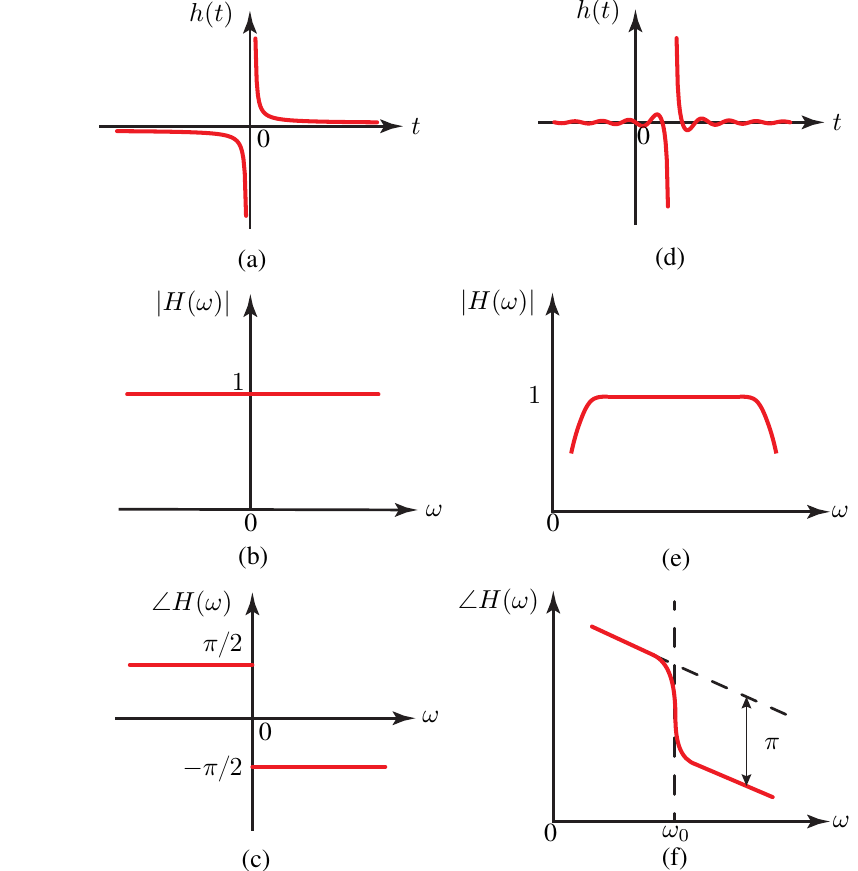}
        %\psfragfig[width=0.98\columnwidth, trim={-1in 0in -0.1in 0in}]{Figs/HTResponse1}{
%        \psfrag{A}[c][c][0.75]{$\pi$}
%        \psfrag{x}[c][c][0.75]{$\omega$}
%        \psfrag{y}[r][r][0.75]{$\angle H(\omega)$}
%        \psfrag{z}[r][r][0.75]{$ |H(\omega)|$}
%        \psfrag{w}[r][r][0.75]{$ h(t)$}
%        \psfrag{m}[c][c][0.75]{$\omega_0$}
%        \psfrag{p}[r][r][0.75]{$\pi/2$}
%        \psfrag{q}[r][r][0.75]{$-\pi/2$}
%        \psfrag{o}[r][r][0.75]{$1$}
%        \psfrag{a}[c][c][0.75]{(a)}
%        \psfrag{b}[c][c][0.75]{(b)}
%        \psfrag{c}[c][c][0.75]{(c)}
%        \psfrag{d}[c][c][0.75]{(d)}
%        \psfrag{e}[c][c][0.75]{(e)}
%        \psfrag{f}[c][c][0.75]{(f)}
%        \psfrag{0}[c][c][0.75]{0}
%        \psfrag{1}[c][c][0.75]{$t$}
%        }
        \caption{Constitutive functions of the mathematical (left) and physical RAP (right) Hilbert transform operating around the center frequency of $\omega_0$. (a),(d)~Impulse response. (b),(e)~Amplitude of the transfer function. (c),(f)~Phase of the transfer function.}
   \label{FIG:HTResp}
\end{figure}

The \emph{Hilbert transformation}, $\mathcal{H}[\cdot]$, is the linear operation depicted in \figref{FIG:HTSche}. It transforms an input signal $x(t)$ into the output signal $y(t)$ by convolving it with the impulse response
\begin{equation}\label{EQ:ht}
h(t)=\frac{1}{\pi t},
\end{equation}
plotted in \figref{FIG:HTResp}(a), as~\cite{Book:Hahn_1996_Hilbert}
\begin{equation}\label{EQ:HT_Time}
  y(t)=\mathcal{H}[{x}(t)]=h(t)*x(t)=\frac{1}{\pi}\ \mathcal{P.V.}\int_{-\infty}^{+\infty}\frac{x(\tau)}{t-\tau}d\tau,
\end{equation}
where $\mathcal{P.V.}$ denotes the Cauchy principal value, accommodating for the fact that $h(t)$ is not integrable at $t=0$.

The Fourier transform of~\eqref{EQ:HT_Time} reads
\begin{equation}
Y(\omega)=H(\omega)X(\omega)
=\mathcal{F}\{\mathcal{H}[x(t)]\},
\end{equation}
where $H(\omega)$ is the transfer function
\begin{equation}\label{EQ:HT_Freq}
  H(\omega)=
  \begin{cases}
  i=e^{+\frac{i\pi}{2}}, & \text{if}~\omega<0,\\
  0,& \text{if}~\omega=0,\\
  -i=e^{-\frac{i\pi}{2}},& \text{if}~\omega>0,
  \end{cases}
\end{equation}
whose magnitude and phase are plotted in Figs.~\ref{FIG:HTResp}(b) and~\ref{FIG:HTResp}(c), respectively.

The all-pass magnitude and step-rotation phase of $H(\omega)$ in~\eqref{EQ:HT_Freq} are mathematically useful but practically noncausal if $t$ is understood as time. In a physical RAP realization, the closest possible magnitude response is the band-pass response shown in \figref{FIG:HTResp}(e) associated with the gradual phase response shown in \figref{FIG:HTResp}(f) and corresponding to the impulse function plotted in \figref{FIG:HTResp}(d). If the signal to process is completely included in the pass-band [\figref{FIG:HTResp}(e)] of transformer, then the magnitude limitation is practically not problematic. The consequence of the phase deviation is more subtle, but also not fundamentally problematic in practice. The (nonzero) asymptotic slopes around $\omega_0$ correspond to simple nondispersive delays, which play no processing role other than shifting the entire processed pulse in time. On the other hand, the progressive phase transition is not prohibitive if the signals of interest do not carry energy in that region.

Note that applying the Hilbert transform twice restores the initial function except for a negative sign. Indeed,
\begin{equation}\label{EQ:recov}
\begin{split}
\mathcal{H}\{\mathcal{H}[x(t)]\}
&=h(t)*[h(t)*x(t)] \\
&=[h(t)*h(t)]*x(t) \\
&=-\delta(t)*x(t)
=-x(t).
\end{split}
\end{equation}
So, once a signal has been processed by a Hilbert transformer, it can be recovered if needed.

\section{Coupler-Resonator based Hilbert Transformer Analysis}\label{SEC：Micro_Imple}

\begin{figure}[h!t]
    \centering
    \includegraphics[width=0.98\columnwidth]{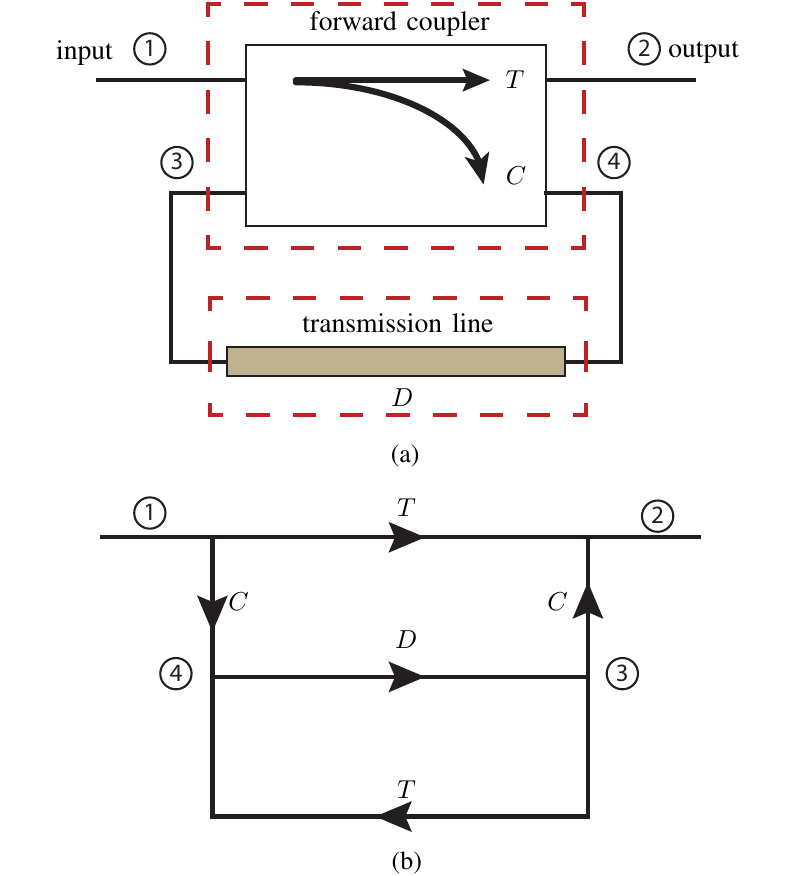}
        %\psfragfig[width=0.9\columnwidth, trim={-0.5in 0in -0.5in 0in}]{Figs/HTScheme}{
%        \psfrag{A}[c][c][0.8]{input}
%        \psfrag{B}[c][c][0.8]{output}
%        \psfrag{c}[c][c][0.8]{forward coupler}
%        \psfrag{t}[c][c][0.8]{transmission line}
%        \psfrag{P}[c][c][0.75]{(a)}
%        \psfrag{Q}[c][c][0.75]{(b)}
%        \psfrag{T}[c][c][0.75]{$T$}
%        \psfrag{S}[c][c][0.75]{$D$}
%        \psfrag{C}[c][c][0.75]{$C$}
 %      }
        \caption{Typical implementation of a RAP Hilbert transformer. (a)~Physical schematics. (b)~Signal flow chart.}
   \label{FIG:Schematic}
\end{figure}

How can one realize a physical component with a transfer function of the type shown in Figs.~\ref{FIG:HTResp}(e) and~\ref{FIG:HTResp}(f)? Such a component should pass all the signal power in the operation frequency band, according to \figref{FIG:HTResp}(e), and exhibit a sharp phase rotation, and hence a group delay [$\tau(\omega)=|\partial\phi(\omega)/\partial\omega|$] peak, at the center frequency $\omega_0$, according to \figref{FIG:HTResp}(f). The first requirement suggests a waveguide and the second a resonant delay element coupled to it. This leads to the structure shown in Fig.~\ref{FIG:Schematic}(a), which is composed of a straight waveguide connecting the input ({\textcircled{\small 1}}) to the output ({\textcircled{\small 2}}) and coupling to a transmission-line loop resonator ({\textcircled{\small 3}}-{\textcircled{\small 4}}-{\textcircled{\small 3}}) with coupled section {\textcircled{\small 3}}-{\textcircled{\small 4}}.

The structure of \figref{FIG:Schematic}(a) may be modeled by the signal flow chart drawn in \figref{FIG:Schematic}(b), where $T$ and $C$, denote the transmission and coupling coefficients of the coupler, respectively, $D$ denotes the transmission coefficient of the transmission-line loop, and where the coupler isolation is assumed to be infinite. Using standard flow-chart rules~\cite{BK:2011_Pozar} yields then the transfer function
\begin{equation}\label{EQ:TransFunc}
  S_{21}=T+\frac{C^2D}{1-TD},
\end{equation}
where $T$, $C$ and $D$ are complex quantities. Assuming that the coupler is passive, power conservation demands
\begin{equation}\label{EQ:Conservation}
|T|^2+|C|^2=1.
\end{equation}
Combining this magnitude relation with the phase relation
\begin{equation}\label{EQ:angle}
\angle C = \angle T - \frac{\pi}{2},
\end{equation}
anticipating the later use of a branch-line coupler [\figref{FIG:CouplerResonator}(b)], leads to complex relation
\begin{equation}\label{EQ:TC}
C^2=T^2-e^{j2\theta},
\end{equation}
where $\theta=\angle T$. Moreover, assuming that the transmission-line loop (minus the part common with the coupler section) is lossless and characterized by the time delay of the  $\tau_0$, we have
\begin{equation}\label{EQ:D}
  D=e^{-j\omega\tau_0}.
\end{equation}

Substituting~\eqref{EQ:TC} and ~\eqref{EQ:D} into ~\eqref{EQ:TransFunc} finally yields the following expression for the transfer function of the Hilbert transformer as a function of the complex quantity $T$\footnote{Later studies will investigate the response of the Hilbert transformer in terms of the coupling, which would suggest to express $S_{21}$ as a function of $C$ rather than $T$. However, it turns out that the latter expression is much more complicated than the former, while offering no specific benefit. For this reason, we have decided to give here $S_{21}$ as a function of $T$, where $T=\sqrt{1-|C|^2}e^{j(\angle C+\pi/2)}$ according to~\eqref{EQ:Conservation} ($|T|=\sqrt{1-|C|^2}$) and \eqref{EQ:angle} ($\angle T=\angle C+\pi/2$).}:
\begin{equation}\label{EQ:Trans}
S_{21}(\omega;T)=\frac{T-e^{-j(\omega_0\tau_0-2\theta)}}{1-Te^{-j\omega\tau_0}},
\end{equation}
from which the system phase and group delay may be computed as
\begin{subequations}\label{EQ:phase_delay}
\begin{equation}\label{EQ:phase}
\phi(\omega;T)=\angle\left\{S_{21}(\omega;T)\right\}
\end{equation}
and
\begin{equation}\label{EQ:delay}
\tau(\omega;T)=-\frac{\partial\phi(\omega;T)}{\partial\omega},
\end{equation}
\end{subequations}
respectively.

\begin{figure}[h!t]
    \centering
    \includegraphics[width=0.9\columnwidth]{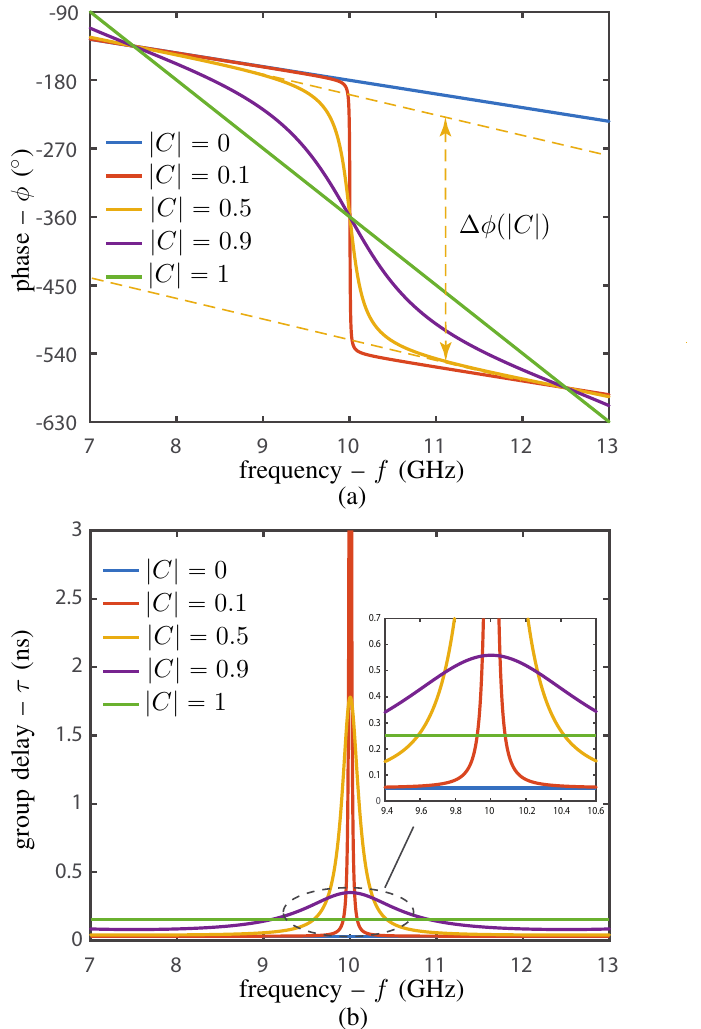}
       % \psfragfig[width=0.9\columnwidth, trim={-0.1in 0in -0.1in 0in}]{Figs/PhaseGrD2}{
%        \psfrag{A}[l][l][0.75]{$2\pi$}
%        \psfrag{B}[l][l][0.75]{$1.5\pi$}
%        \psfrag{H}[l][l][0.75]{$\pi$}
%        \psfrag{C}[l][l][0.65]{$|C|=0$}
%        \psfrag{D}[l][l][0.65]{$|C|=0.1$}
%        \psfrag{E}[l][l][0.65]{$|C|=0.5$}
%        \psfrag{F}[l][l][0.65]{$|C|=0.9$}
%        \psfrag{G}[l][l][0.65]{$|C|=1$}
%        \psfrag{P}[l][l][0.75]{$\Delta\phi(|C|)$}
%        \psfrag{X}[c][c][0.75]{frequency -- $f$ (GHz)}
%        \psfrag{Y}[c][c][0.75]{phase -- $\phi$ ($^\circ$)}
%        \psfrag{Z}[c][c][0.75]{group delay -- $\tau$ (ns)}
%        \psfrag{a}[c][c][0.75]{(a)}
%        \psfrag{b}[c][c][0.75]{(b)}
%        }
        \caption{Frequency response of the RAP Hilbert transformer in \figref{FIG:Schematic} for different coupling coefficient magnitudes, $|C|$, and $f_0=\omega_0/(2\pi)=10$~GHz and $\tau_0=3\pi/\omega_0=0.15$~ns. (a)~Phase [Eq.~\eqref{EQ:phase}]. (b)~Group delay [Eq.~\eqref{EQ:delay}].}
   \label{FIG:Coupling}
\end{figure}

If the coupled part of the loop is $\lambda/2$-long, or $\theta=\pi$, as will be the case with the final (cascaded double) branch-line coupler (\figref{FIG:Layout}), the shortest possible length of the uncoupled part of the transmission-line loop is $3\lambda/2$, or $\tau_0=3\pi/\omega_0$, corresponding to a $2\lambda$ resonant loop. Figures~\ref{FIG:Coupling}(a) and~\ref{FIG:Coupling}(b) plot the phase and group delay obtained by~\eqref{EQ:phase} and~\eqref{EQ:delay} for this scenario with center frequency $f_0=\omega_0/(2\pi)=10$~GHz. These results will be commented in Sec.~\ref{sec:transf_char} and physically explained in Sec.~\ref{SEC:PhysExp}.

\section{Transformer Characterization}\label{sec:transf_char}

As announced in \figref{FIG:HTResp}(f) and verified in \figref{FIG:Coupling}(a), the phase response of a RAP Hilbert transformer significantly departs from that of the ideal mathematical Hilbert transformer, plotted in \figref{FIG:HTResp}(c): specifically, 1)~the \emph{asymptotic slopes} for $\omega\ll\omega_0$ and $\omega\gg\omega_0$ are \emph{nonzero}, and 2)~the \emph{transition bandwidth} between the asymptotic slopes is \emph{nonzero}.

\begin{figure}[h!]
    \centering
    \includegraphics[width=0.9\columnwidth]{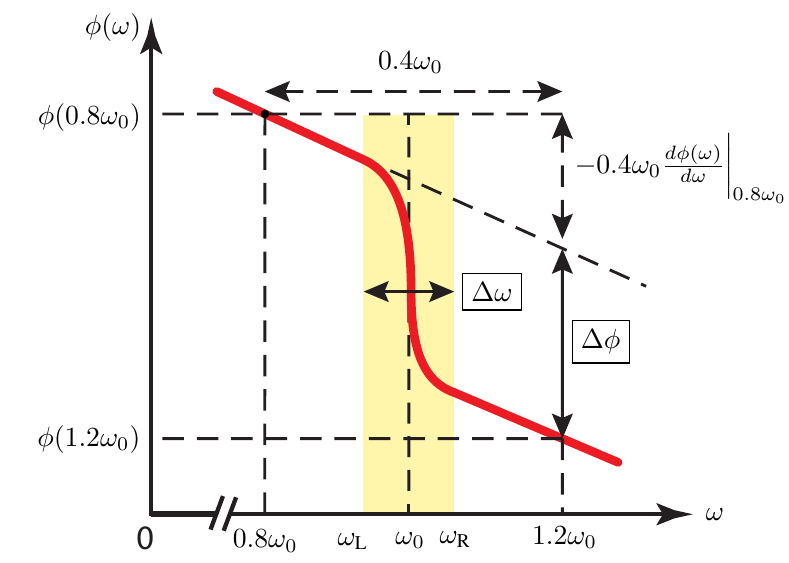}
        %\psfragfig*[width=0.9\columnwidth, trim={-0.5in 0in -0.3in 0in}]{Figs/RotatedPhase}{
%        \psfrag{A}[c][c][0.8]{$\boxed{\Delta\phi}$}
%        \psfrag{D}[c][c][0.8]{$\boxed{\Delta\omega}$}
%        \psfrag{B}[l][l][0.8]{$-0.4\omega_0\frac{d\phi(\omega)}{d\omega}\bigg|_{0.8\omega_0}$}
%        \psfrag{C}[c][c][0.8]{$0.4\omega_0$}
%        \psfrag{a}[c][c][0.8]{$0.8\omega_0$}
%        \psfrag{b}[c][c][0.8]{$1.2\omega_0$}
%        \psfrag{m}[c][c][0.8]{$\omega_0$}
%        \psfrag{c}[r][r][0.8]{$\phi(1.2\omega_0)$}
%        \psfrag{d}[r][r][0.8]{$\phi(0.8\omega_0)$}
%        \psfrag{e}[r][r][0.8]{$\omega_\text{L}$}
%        \psfrag{f}[r][r][0.8]{$\omega_\text{R}$}
%        \psfrag{x}[c][c][0.8]{$\omega$}
%        \psfrag{y}[c][c][0.8]{$\phi(\omega)$}
%       }
        \caption{Characterization of the RAP Hilbert transformer in \figref{FIG:Schematic}, corresponding to \figref{FIG:HTResp}(f): rotated phase, $\Delta\phi$, and transition bandwidth, $\Delta\omega$.}
   \label{FIG:RotPhas}
\end{figure}

In order to quantitatively characterize physical Hilbert transformers in terms of these two aspects, we define here related quantities, with the help of \figref{FIG:RotPhas}.

We define the \emph{rotated phase} as the phase difference between the tangents to the phase curves at $80\%$ and $120\%$ of $\omega_0$ ($40\%$ centered bandwidth), namely
\begin{equation}\label{EQ:Phase_Rotation}
\Delta\phi(|C|)=\phi(0.8\omega_0)-\phi(1.2\omega_0)+0.4\omega_0\frac{d\phi(\omega)}{d\omega}\bigg|_{0.8\omega_0},
\end{equation}
which depends on the coupling level, $|C|$, according to \figref{FIG:Coupling}.

Moreover, we define the \emph{transition bandwidth}, which is also a function of $|C|$, as the symmetric band delimited by the frequencies $\omega_\text{L}$ and $\omega_\text{R}$ whose slopes depart by a factor $\alpha$ from the asymptotic slopes at $80\%$ and $120\%$ of $\omega_0$, namely
\begin{subequations}\label{EQ:Bandwidth Transition}
\begin{equation}
\Delta\omega(|C|)
=\omega_\text{R}-\omega_\text{L}
=2(\omega_0-\omega_\text{L})
=2(\omega_\text{R}-\omega_0),
\end{equation}
where $\omega_\text{L}$ and $\omega_\text{R}$ are defined via
\begin{equation}
\dfrac{\dfrac{d\phi(\omega)}{d\omega}\biggl|_{\omega_\text{L}}
-\dfrac{d\phi(\omega)}{d\omega}\biggl|_{0.8\omega_0}}{\dfrac{d\phi(\omega)}{d\omega}\biggl|_{0.8\omega_0}}
=\dfrac{\dfrac{d\phi(\omega)}{d\omega}\biggl|_{\omega_\text{R}}
-\dfrac{d\phi(\omega)}{d\omega}\biggl|_{1.2\omega_0}}{\dfrac{d\phi(\omega)}{d\omega}\biggl|_{1.2\omega_0}}
=\alpha.
\end{equation}
\end{subequations}

Note that the definitions~\eqref{EQ:Phase_Rotation} and~\eqref{EQ:Bandwidth Transition} are not valid at the limit cases $C=0$ and $|C|=1$, as may be understood by inspecting \figref{FIG:Coupling}(a).

\begin{figure}[h!t]
    \centering
    \includegraphics[width=0.9\columnwidth]{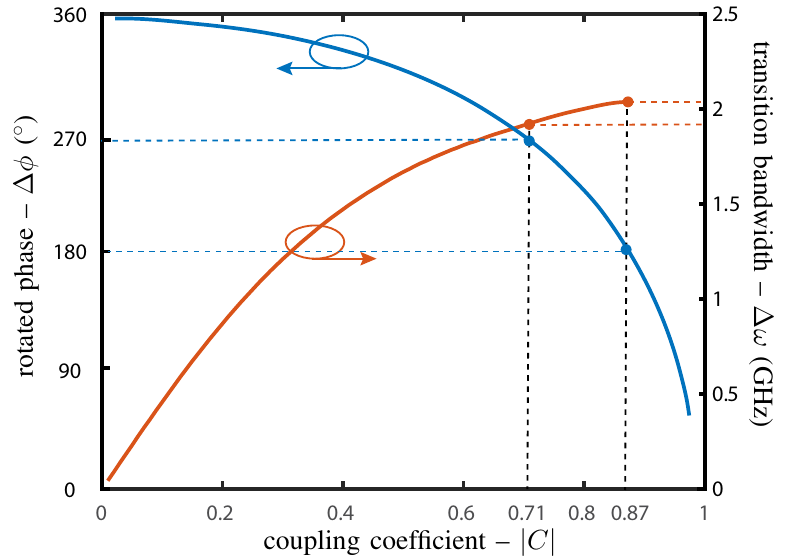}
        %\psfragfig[width=0.9\columnwidth, trim={-0.1in 0in -0.1in 0in}]{Figs/tradoff}{
%        \psfrag{A}[c][c][0.8]{coupling coefficient -- $|C|$}
%        \psfrag{B}[c][c][0.8]{rotated phase -- $\Delta\phi$ ($^\circ$)}
%        \psfrag{C}[c][c][0.8]{transition bandwidth -- $\Delta\omega$ (GHz)}
%        }
        \caption{Rotated phase [Eq.~\eqref{EQ:Phase_Rotation}] and transition bandwidth [Eq.~\eqref{EQ:Bandwidth Transition} with $\alpha=0.35$] versus coupling magnitude $|C|$, for the same parameters as in \figref{FIG:Coupling}.}
   \label{FIG:tradeoff}
\end{figure}

Figure~\ref{FIG:tradeoff} plots the rotated phase and transition bandwidth versus coupling level. It shows that the rotated phase ($\Delta\phi$) is \emph{inversely proportional} to the coupling level ($|C|$) while the transition bandwidth ($\Delta\omega$), being inversely proportional to the rotated phase, is \emph{proportional} to the coupling level. This may be \emph{mathematically} explained in terms of the group delay, plotted in \figref{FIG:Coupling}(b): since the phase is the integral of the group delay [Eq.~\eqref{EQ:delay}], the transition bandwidth (resp. rotated phase) is necessarily proportional (resp. inversely proportional) to the group delay bandwidth, and hence inversely proportional (resp. proportional) with the coupling level. The physical explanation of the observed group delay response will be provided in~\ref{SEC:PhysExp}.

\section{Microwave Implementation}\label{sec:micr_impl}
Figure~\ref{FIG:tradeoff} shows that the $180^\circ$ rotated phase needed for the RAP Hilbert transformer [\figref{FIG:HTResp}(f)] would require a coupling coefficient of $|C|=0.87$. Such a high coupling level is difficult to attain in microwave couplers. Moreover, also according to \figref{FIG:tradeoff}, it would lead to a relatively large -- perhaps undesirably large - transition bandwidth.

To avoid these two issues, we propose to \emph{cascade two identical couplers of lower coupling} (and hence also of smaller transition bandwidth) to realize the required total $180^\circ$ rotated phase. Specifically, we choose a single-coupler rotated phase of $270^\circ$, corresponding to a total rotated phase of $2\times270^\circ=540^\circ\triangleq 180^\circ$. This, according to \figref{FIG:tradeoff} corresponds to a coupling coefficient of $|C|=0.71$, associated with a transition bandwidth that is narrower than that associated with $|C|=0.87$.

The largest coupling bandwidths at microwaves are provided by coupled-line couplers. However, these couplers are restricted to low coupling levels, typically substantially smaller than 3-dB. Therefore, we decide to use here a \emph{branch-line coupler}~\cite{BK:2011_Pozar}, whose coupling level of $|C|=1/\sqrt{2}=0.707\approx0.71$ is the most commonly used in practice.

\begin{figure}[h!t]
    \centering
    \includegraphics[width=1\columnwidth]{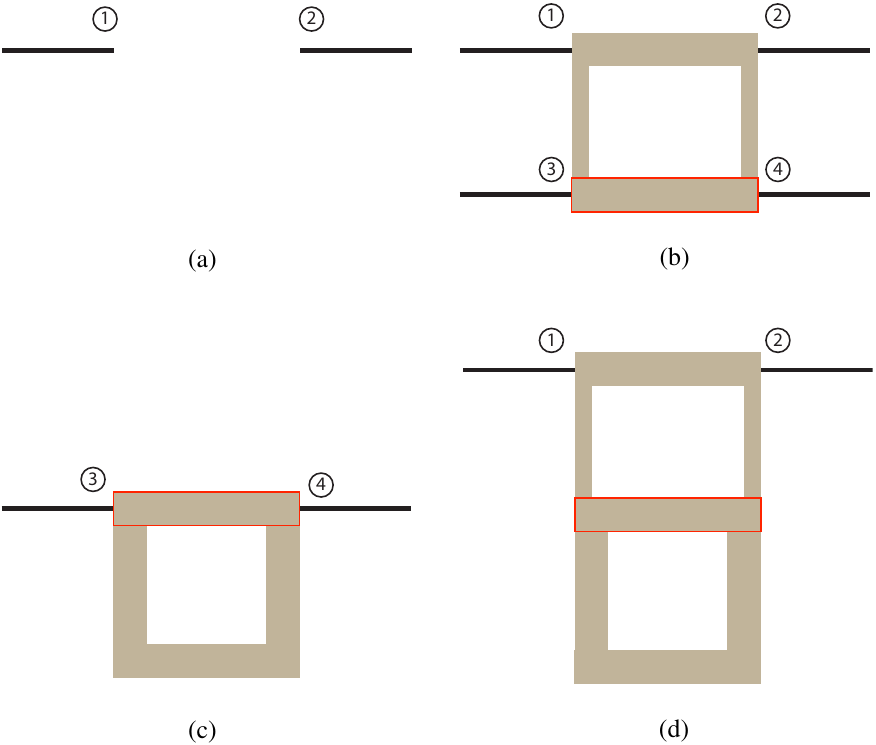}
        %\psfragfig[width=1\columnwidth, trim={0in 0in 0in 0in}]{Figs/coupler_resonator}{
%        \psfrag{a}[c][c][0.75]{(a)}
%        \psfrag{b}[c][c][0.75]{(b)}
%        \psfrag{c}[c][c][0.75]{(c)}
%        \psfrag{d}[c][c][0.75]{(d)}
%       }
        \caption{Assembly of a unit of the proposed microwave RAP Hilbert transformer. (a)~Input and output ports. (b)~Branch-line coupler. (c)~Loop resonator. (d)~Assembly. The transmission-line sections within the red boxes in (b) and (c) are merged together in (d).}
   \label{FIG:CouplerResonator}
\end{figure}

Figure~\ref{FIG:CouplerResonator} shows the corresponding layout composition for a \emph{single unit}, with the input-output connections, the branch-line coupler itself, the loop resonator and the assembly of the last two given in Figs.~\ref{FIG:Schematic}(a), \ref{FIG:Schematic}(b), \ref{FIG:Schematic}(c) and~\ref{FIG:Schematic}(d), respectively.

\begin{figure}[h!t]
    \centering
    \includegraphics[width=1\columnwidth]{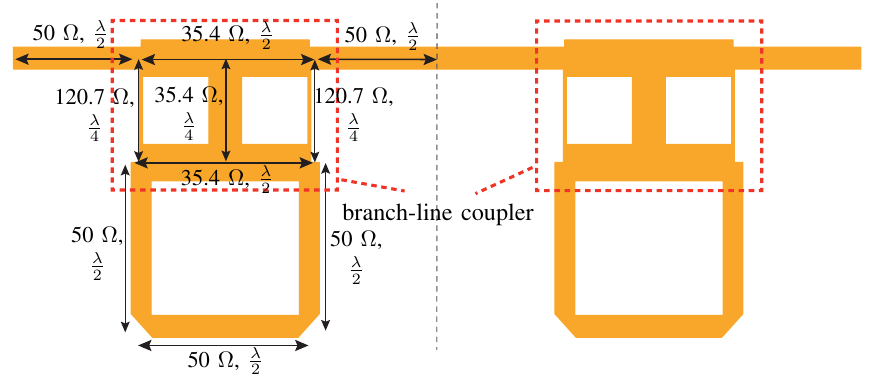}
        %\psfragfig*[width=1\columnwidth, trim={-0.1in 0in -0.1in 0in}]{Figs/Layout}{
%        \psfrag{I}[l][l][0.7]{input}
%        \psfrag{o}[r][r][0.7]{output)}
%        \psfrag{a}[r][r][0.65]{\shortstack{120.7 $\Omega$,\\ $\frac{\lambda}{4}$}}
%        \psfrag{b}[c][c][0.65]{35.4 $\Omega$,\ $\frac{\lambda}{2}$}
%        \psfrag{c}[c][c][0.65]{\shortstack{50 $\Omega$,\\ $\frac{\lambda}{2}$}}
%        \psfrag{d}[c][c][0.65]{50 $\Omega$,\ $\frac{\lambda}{2}$}
%        \psfrag{e}[r][r][0.65]{\shortstack{35.4 $\Omega$,\\ $\frac{\lambda}{4}$}}
%        \psfrag{A}[c][c][0.7]{branch-line coupler}
 %       }
        \caption{Layout of the proposed RAP microwave Hilbert transformer, composed of two cascaded units of $270^\circ$ rotated phase including each a double-section branch-line coupler.}
   \label{FIG:Layout}
\end{figure}

As explained above, the single-unit assembly in \figref{FIG:CouplerResonator}(d) will be designed to provide a $270^\circ$ rotated phase, and two such units will be cascaded to achieve the required $2\times270^\circ=540^\circ\triangleq 180^\circ$ rotated phase. However, the bandwidth of a branch-line coupler is typically in the order of $15\%$~\cite{BK:2011_Pozar} while the transition bandwidth for the selected coupling level ($|C|=0.71$) is of around $20\%$, according \figref{FIG:tradeoff}. Therefore, we replace the single-section branch-line coupler by a double-section branch-line coupler~\cite{BK:2011_Pozar}. Figure~\ref{FIG:Layout} the layout of the overall proposed RAP microwave Hilbert transform with its two cascaded units including each a double-section branch-line coupler.

\section{Physical Explanation}\label{SEC:PhysExp}
\begin{figure}[h!]
  \centering
  \includegraphics[width=0.85\columnwidth]{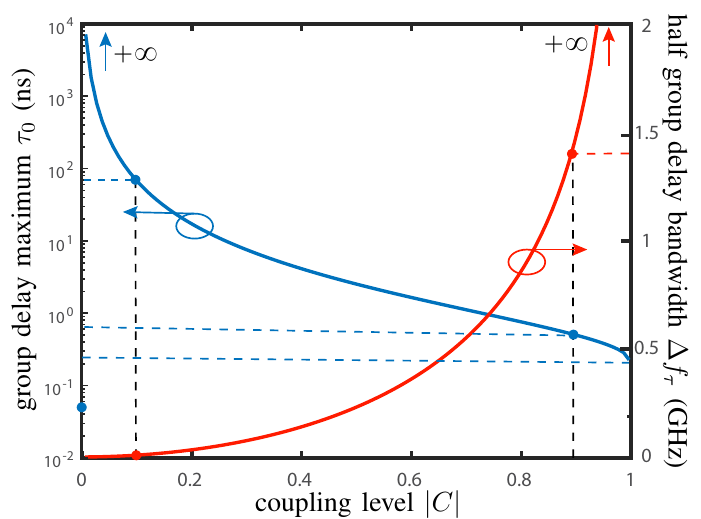}
        %\psfragfig[width=0.85\columnwidth, trim={-0.1in 0in -0.1in -0.1in}]{Figs/GroupDelayC}{
%        \psfrag{A}[c][c][0.8]{coupling level $|C|$}
%        \psfrag{B}[c][c][0.8]{group delay maximum $\tau_0$ (ns)}
%        \psfrag{C}[c][c][0.8]{half group delay bandwidth $\Delta f_{\tau_0/2}$ (GHz)}
%        \psfrag{D}[c][c][0.8]{$+\infty$}
%        \psfrag{E}[c][c][0.8]{$+\infty$}
%        }
  \caption{Group delay maximum, $\tau(\omega_0)=\tau_0$ (log scale), and half group delay bandwidth, $\Delta f_{\tau_0/2}$ versus coupling level $|C|$.}\label{FIG:tau0_vs_C}
\end{figure}

Section~\ref{SEC：Micro_Imple} derived the transfer function and plotted the phase and group delay responses of the RAP Hilbert transformer, while Sec.~\ref{sec:transf_char} characterized the corresponding rotated phase and transition bandwidth responses in term of the coupling level, pointing out how the device response is related to the group delay versus coupling response. This section will provide the physical explanation of this group delay response -- and specifically the decrease of the group delay peak $\tau(\omega_0)=\tau_0$ versus coupling level $|C|$, plotted in \figref{FIG:tau0_vs_C} -- for microwave implementations of the type proposed in Sec.~\ref{sec:transf_char}, using both a steady-state regime and a transient regime perspectives.

\subsection{Steady-State Regime}

Table~\ref{TAB:Explanation} compares the steady-state harmonic regime operation of the RAP Hilbert transformer at the four different coupling levels indicated in \figref{FIG:tau0_vs_C}. Let us consider these four cases one by one.

\begin{table}[h!]
  \centering
  \caption{Steady-state explanation for the group delay response of the RAP Hilbert transformer.}\label{TAB:Explanation}
\begin{tabular}{c|c|c}

  \ &Case 1:~$|C|=0$ & Case 2:~$|C|=0.1 $\\
  \hline
  \begin{minipage}{8mm} wave \\ flow \\~\\~\\ \end{minipage}&\includegraphics[width=0.35\columnwidth]{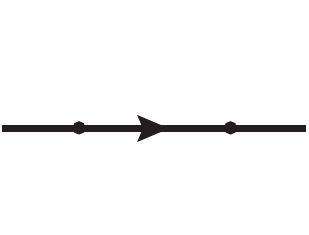}%\psfragfig[width=0.35\columnwidth, trim={0in 0in 0in -0.2in}]{Figs/Explanation_CaseA_1}
    \vspace{-0.5cm}&
  \includegraphics[width=0.35\columnwidth]{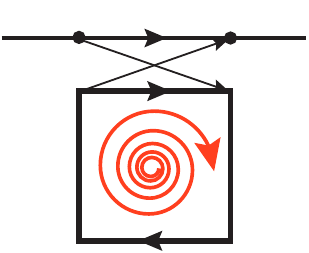}\\\vspace{-0.5cm}
  \begin{minipage}{8mm} group \\ delay \\~\\~\\ \end{minipage}&\includegraphics[width=0.35\columnwidth]{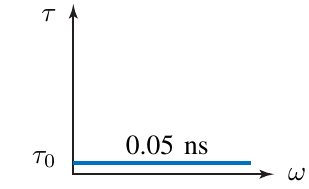}&
  \includegraphics[width=0.35\columnwidth]{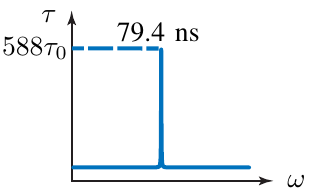}\\
  \hline
  \ &Case 3:~$|C|=0.9$& Case 4:~$|C|=1$\\
  \hline
  \begin{minipage}{8mm} wave \\ flow \\~\\~\\ \end{minipage}&\includegraphics[width=0.35\columnwidth]{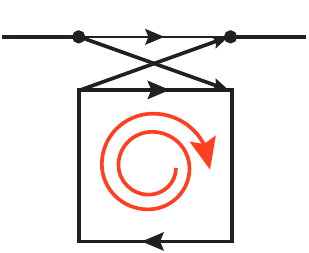}\vspace{-0.5cm}&
  \includegraphics[width=0.35\columnwidth]{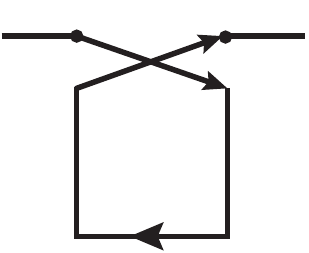}\\\vspace{-0.5cm}
  \begin{minipage}{8mm} group \\ delay \\~\\~\\ \end{minipage}&\includegraphics[width=0.35\columnwidth]{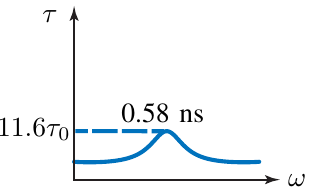}&
  \includegraphics[width=0.35\columnwidth]{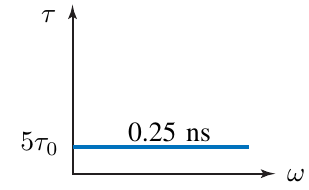}
\end{tabular}
\end{table}
\vspace{5mm}

\subsubsection{$C=0$} This is a very particular case, where the resonant loop brutally becomes invisible (or infinitely far), so that the transformer structure degenerates into a simple straight transmission line. As a result, and according to~\eqref{EQ:TransFunc}, $S_{21}(C=0)=T$, and hence the group delay is $\tau_0=(\lambda/2)/(\lambda f_0)=1/(2f_0)=0.05$~ns (constant). The bandwidth is naturally infinite, assuming an ideal transmission line.

\subsubsection{$|C|=0.1$} As soon as $C\neq 0$, the system recovers its coupled resonator, and therefore behaves completely differently. Figure~\ref{FIG:tau0_vs_C} shows that $\lim_{|C|\rightarrow 0}\tau_0=\infty$. Why is this the case? If $|C|$ is nonzero but very small, a small amount of energy per harmonic cycle ($1\%$ for $|C|=0.1$) couples into the resonant loop, but an even smaller amount of energy ($1\%$ of the remainder for $|C|=0.1$) couples out at each turn, so that most of the coupled energy loops in the resonator for a very long time, hence yielding a very high group delay ($1158\tau_0=79.4$~ns for $|C|=0.1$). Moreover, the bandwidth is very narrow because the resonator is seen by the straight transmission line as a lossy load with loss proportional to coupling, leading to an effective quality factor inversely proportional to the coupling level, i.e. $Q\,\propto\,1/\Delta\omega\,\propto\,1/|C|$.

\subsubsection{$|C|=0.9$} At large coupling levels, most of the energy ($81\%$ for $|C|=0.9$) of each harmonic cycle is coupled into the resonant loop, but a lot of energy is still coupled out of it ($66\%$ after the first turn for $|C|=0.9$), so that most of the energy is mostly evacuated after a small number of turns in the loop, yielding a much smaller group delay ($11.6\tau_0=0.58$~ns for $|C|=0.9$). At the same time, the quality factor of the system, $Q\,\propto\,1/|C|$, is now much smaller, and therefore the group delay bandwidth is strongly increased.

\subsubsection{$|C|=1$} In contrast to $C=0$ with $|C|\rightarrow 0$, the case $|C|=1$ is in continuity with $|C|\rightarrow 1$. As $|C|\rightarrow 1$, the effect of the straight line gradually diminishes and exactly disappears at$|C|=1$, where the harmonic cycle energy performs exactly one turn in the loop, corresponding to a constant delay of $5\tau_0=0.25$~ns, again with infinite bandwidth.

\subsection{Transient Regime}
The steady-state harmonic regime is most informative, as it corresponds to the operation regime of the RAP Hilbert transformer. However, it is somewhat difficult to apprehend the notion of group delay for a continuous wave, since such a wave has neither beginning nor end. Therefore, we offer here an alternative perspective, the transient harmonic regime perpspective, where a continuous wave of frequency $\omega_0$ is injected into the system at time $t=0$.

\begin{figure}[h!]
    \centering
    \includegraphics[width=0.8\columnwidth]{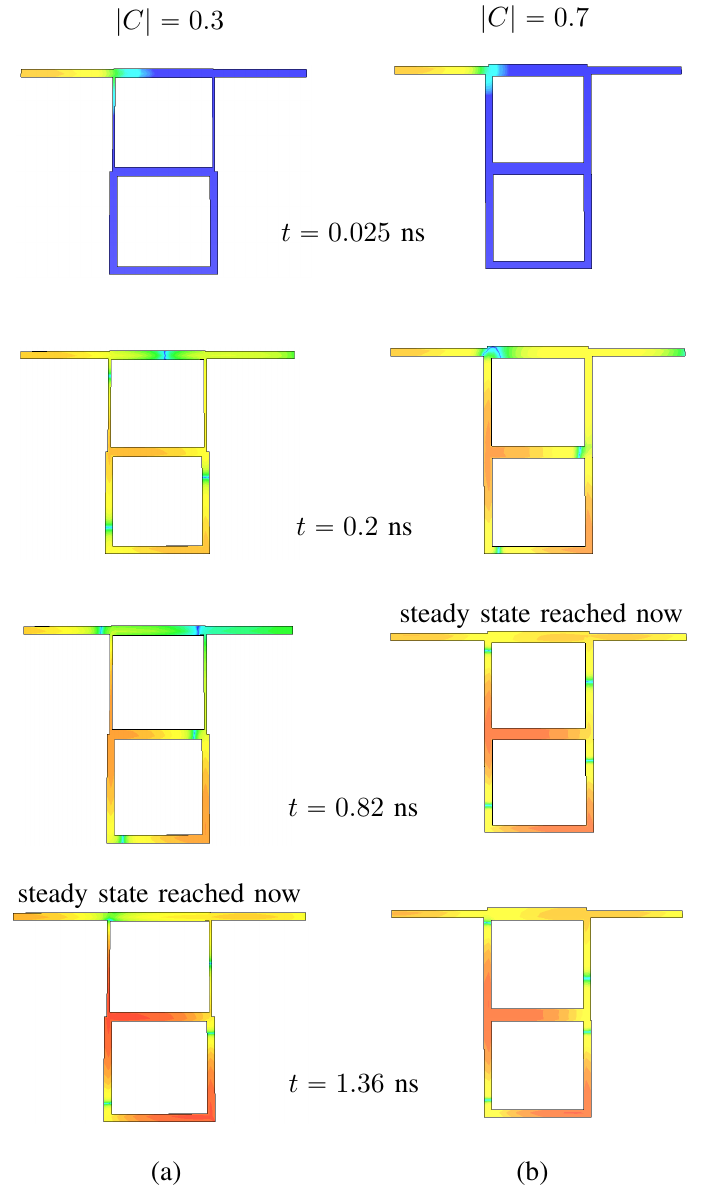}
       % \psfragfig[width=0.8\columnwidth, trim={-0.1in 0in -0.1in -0.1in}]{Figs/field}{
%        \psfrag{a}[c][c][0.8]{$t=0.025$~ns}
%        \psfrag{b}[c][c][0.8]{$t=0.2$~ns}
%        \psfrag{c}[c][c][0.8]{$t=0.82$~ns}
%        \psfrag{d}[c][c][0.8]{$t=1.36$~ns}
%        \psfrag{A}[c][c][0.8]{(a)}
%        \psfrag{B}[c][c][0.8]{(b)}
%        \psfrag{C}[c][c][0.8]{$|C|=0.7$}
%        \psfrag{D}[c][c][0.8]{$|C|=0.3$}
%        \psfrag{E}[c][c][0.8]{steady state reached now}
%        }
        \caption{Transient full-wave explanation for the group delay in a RAP Hilbert transformer. (a)~Small coupling coefficient, $|C|=0.3$, corresponding to large group delay of $\tau_\text{sim.}=0.83$ ns. (b)~Large coupling coefficient, $|C|=0.7$, corresponding to small group delay of $\tau_\text{sim.}=1.30$ ns.}
   \label{FIG:Transient}
\end{figure}

Figure~\ref{FIG:Transient} shows the full-wave simulated evolution of this wave into a simplified (single section with mono-section branch-line couplers) version -- for easier visualization -- of the microwave Hilbert transformer in \figref{FIG:Layout} for two different coupling levels. In both cases, one sees that the energy first gradually penetrates into the structure, loads the resonant loop, and finally reaches the steady-state regime. However, it is observed -- particularly by inspecting the output branch of the structure --- that the steady-state regime is reached later in the low-coupling transformer and earlier in the high-coupling transformer. Moreover, the time taken by the harmonic wave to reach the steady state is found to be close to the simulated group delay, taken here as the maximal delay ($\tau_0$). This time corresponds indeed to the transmission time of a harmonic-wave ``packet'' within the observation time, and is hence an an excellent proxy for the group delay.

\section{Experiment Demonstration}
\begin{figure}[h!t]
    \centering
    \includegraphics[width=0.8\columnwidth]{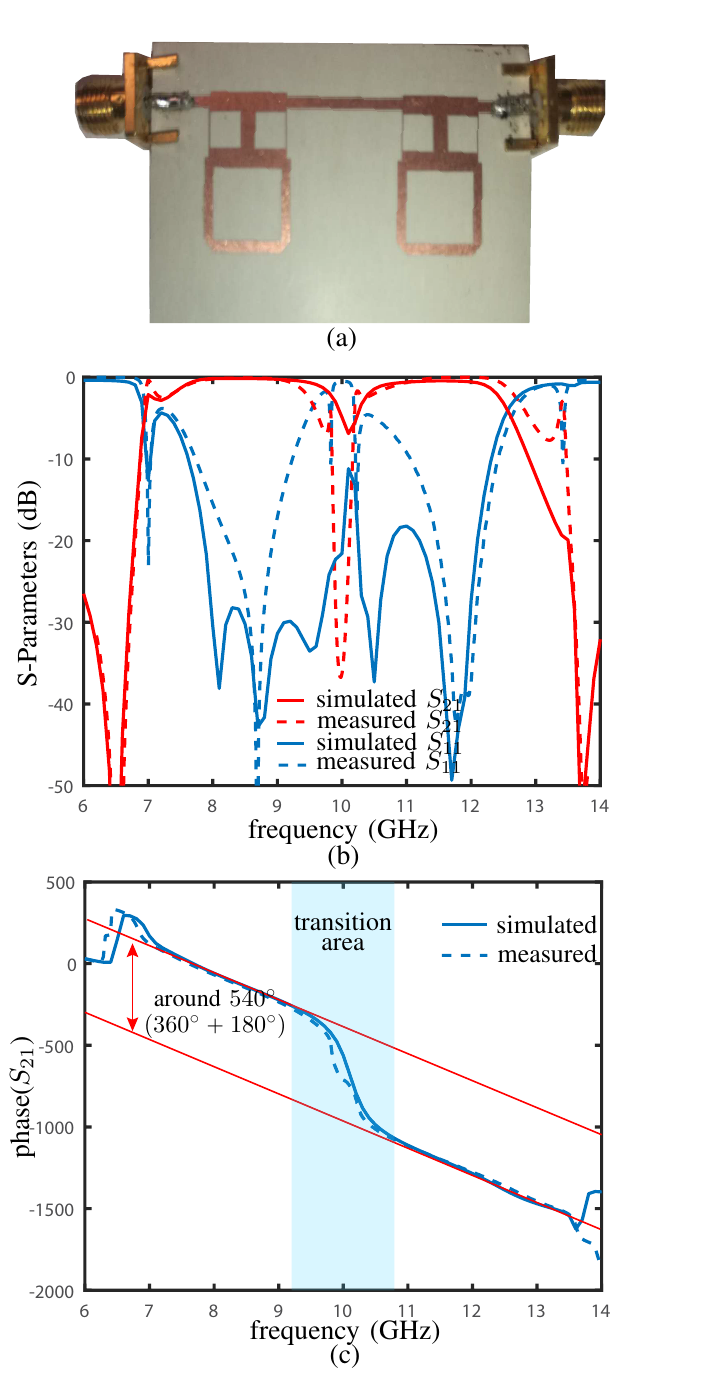}
        %\psfragfig[width=0.8\columnwidth, trim={-0.1in 0in 0in 0in}]{Figs/HT_Exp}{
%        \psfrag{x}[c][c][0.8]{(a)}
%        \psfrag{y}[c][c][0.8]{(b)}
%        \psfrag{z}[c][c][0.8]{(c)}
%        \psfrag{a}[l][l][0.75]{simulated $S_{21}$}
%        \psfrag{b}[l][l][0.75]{measured $S_{21}$}
%        \psfrag{c}[l][l][0.75]{simulated $S_{11}$}
%        \psfrag{d}[l][l][0.75]{measured $S_{11}$}
%        \psfrag{e}[l][l][0.75]{simulated}
%        \psfrag{f}[l][l][0.75]{measured}
%        \psfrag{g}[l][l][0.7]{\shortstack{around $540^\circ$ \\ $(360^\circ+180^\circ)$}}
%        \psfrag{h}[c][c][0.75]{\shortstack{transition \\ area}}
%        \psfrag{A}[c][c][0.8]{frequency (GHz)}
%        \psfrag{B}[c][c][0.8]{S-Parameters (dB)}
%        \psfrag{C}[c][c][0.8]{phase($S_{21}$)}
%        }
        \caption{Experiment demonstration of proposed RAP Hilbert transformer. (a)~Prototype. (b)~Simulated and measured S-parameters. (c)~Phase response.}
   \label{FIG:HT_Exp}
\end{figure}

Figure~\ref{FIG:HT_Exp}(a) shows a fabricated RAP Hilbert transformer prototype with the design and dimensions given in \figref{FIG:Layout}. The corresponding magnitude and phase responses are plotted in Figs.~\ref{FIG:HT_Exp}(b) and~\ref{FIG:HT_Exp}(c), respectively. The simulated and measured results are in fair agreement with each other. The passband extends from 7.8~GHz to 12.2~GHz, except for a notch at the center frequency, 10~GHz. This notch is due to the losses of the material and microstrip line radiation, which are naturally maximal at the frequency, where energy is stored in the structure for the longest time (group delay peak), i.e. the resonance frequency. However, if no energy is injected in its frequency range, this notch does not pose any practical problem.

\section{Applications}
In order to illustrate the usefulness of the proposed microwave RAP Hilbert transformer, we present here three applications of it. In each case, the results have been obtained by applying the scattering matrix of the experimental prototype of \figref{FIG:HT_Exp} to the test signals.

\subsection{Edge Detection}
The Hilbert-transformer is essentially an \emph{edge detector}, and can thus be used for instance  to increase contrast in image processing or enhance the signal-to-noise ratio in differential-modulation communication schemes.

The detection operation can be mathematically demonstrated by convolving the input rectangular pulse
\begin{equation}\label{EQ:Rect}
x(t)=\Pi(t)=
\begin{cases}
  1,& \text{if}~|t| \leq 1,\\
  0,& \text{if}~|t| > 1
\end{cases}
\end{equation}
with the impulse function $h(t)$ in~\eqref{EQ:ht}, according to the definition~\eqref{EQ:HT_Time}, which yields
\begin{equation}\label{EQ:Resp_of_Rect}
\begin{aligned}
  y(t)=&\frac{1}{\pi}\mathcal{P.V.}\int_{-\infty}^{+\infty}\Pi(\tau)\frac{1}{t-\tau}d\tau
      =\frac{1}{\pi}\mathcal{P.V.}\int_{-1}^{1}\frac{1}{t-\tau}d\tau \\
      =&\frac{1}{\pi}\mathcal{P.V.}\int_{t+1}^{t-1}\frac{1}{u}d(-u)\quad(\text{substitution }  u=t-\tau)\\
      =&-\frac{1}{\pi}\ln\left|u\right|\bigg|_{t+1}^{t-1}
      =\frac{1}{\pi}\ln\left|\frac{t+1}{t-1}\right|
      =\mathcal{H}[\Pi(t)].
\end{aligned}
\end{equation}
Figure~\eqref{EQ:Resp_of_Rect} plots both~\eqref{EQ:Rect} and~\eqref{EQ:Resp_of_Rect}, and clearly shows the announced edge detection effect, associated with the poles of~\eqref{EQ:Resp_of_Rect} at $t=\pm1$.
\begin{figure}[h!]
    \centering
    \includegraphics[width=0.78\columnwidth]{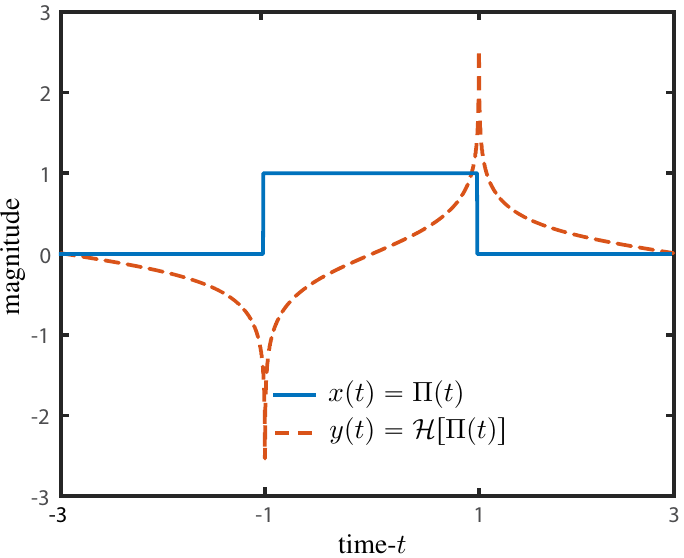}
       % \psfragfig*[width=0.78\columnwidth, trim={0in 0in 0in 0in}]{Figs/Ideal_Hilbert_Square}{
%        \psfrag{a}[c][c][0.8]{time-$t$}
%        \psfrag{b}[c][c][0.8]{magnitude}
%        \psfrag{A}[l][l][0.8]{$x(t)=\Pi(t)$}
%        \psfrag{B}[l][l][0.8]{$y(t)=\mathcal{H}[\Pi(t)]$}
%        }
        \caption{Hilbert transform of a square pulse [Eqs.~\eqref{EQ:Rect} and~\eqref{EQ:Resp_of_Rect}].}
   \label{FIG:Edge_Det_Ideal}
\end{figure}

\begin{figure}[h!]
    \centering
    \includegraphics[width=1\columnwidth]{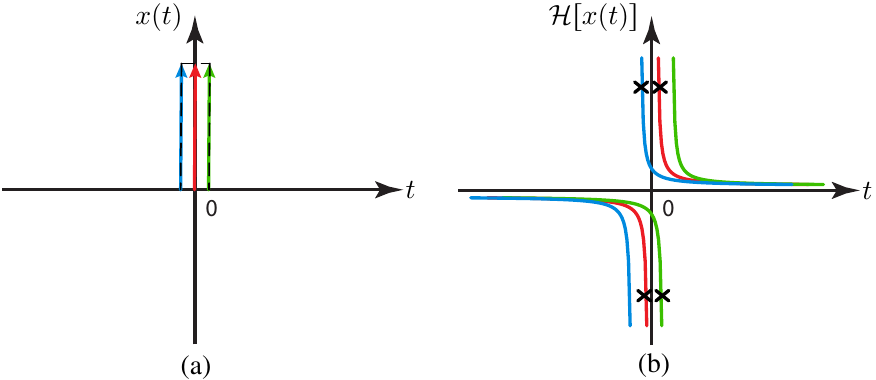}
        %\psfragfig[width=1\columnwidth, trim={0in 0in 0in 0in}]{Figs/ExpSquare}{
%        \psfrag{a}[c][c][0.8]{(a)}
%        \psfrag{b}[c][c][0.8]{(b)}
%        \psfrag{w}[r][r][0.8]{$x(t)$}
%        \psfrag{x}[r][r][0.8]{$\mathcal{H}[x(t)]$}
%        \psfrag{1}[l][l][0.8]{$t$}
%        }
        \caption{Intuitively explanation of the square pulse edge detection in \figref{FIG:Edge_Det_Ideal}. (a)~Input signal rectangular pulse $x(t)=\Pi(t)$ expressed as a (continuous) superposition of Dirac functions. (b)~Formation of the Hilbert transform of (a) using the fact that $y(t)=h(t)$ [Eq.~\eqref{EQ:ht}] for $x(t)=\delta(t)$.}
   \label{FIG:ExpSquare}
\end{figure}

Figure~\ref{FIG:ExpSquare} offers an intuitive explanation for the edge detection just mathematically demonstrated. The pulse function $x(t)=\Pi(t)$ may be constructed as a continuous superposition of Dirac functions across the duration of the pulse, as suggested in \figref{FIG:ExpSquare}(a), since $\Pi(t)=\Pi(t)*\delta(t)=\int_{-1}^{1}\delta(t-t')dt'$. But one knows from~\eqref{EQ:HT_Time} that $y(t)=h(t)$ [Eq.~\eqref{EQ:ht}] for $x(t)=\delta(t)$. Therefore, by superposition (linearity), the response $y(t)$ is the (continuous) sum of the functions $h(t-t')$ with $t'\in[-1,1]$. Since $h(t-t')$ is odd, the signal within the range vanishes due to mutually canceling opposite branches of $h$, and only the edge branches remain, as shown in \figref{FIG:ExpSquare}(b), consistently with \figref{FIG:Edge_Det_Ideal}.

\begin{figure}[h!]
    \centering
    \includegraphics[width=0.8\columnwidth]{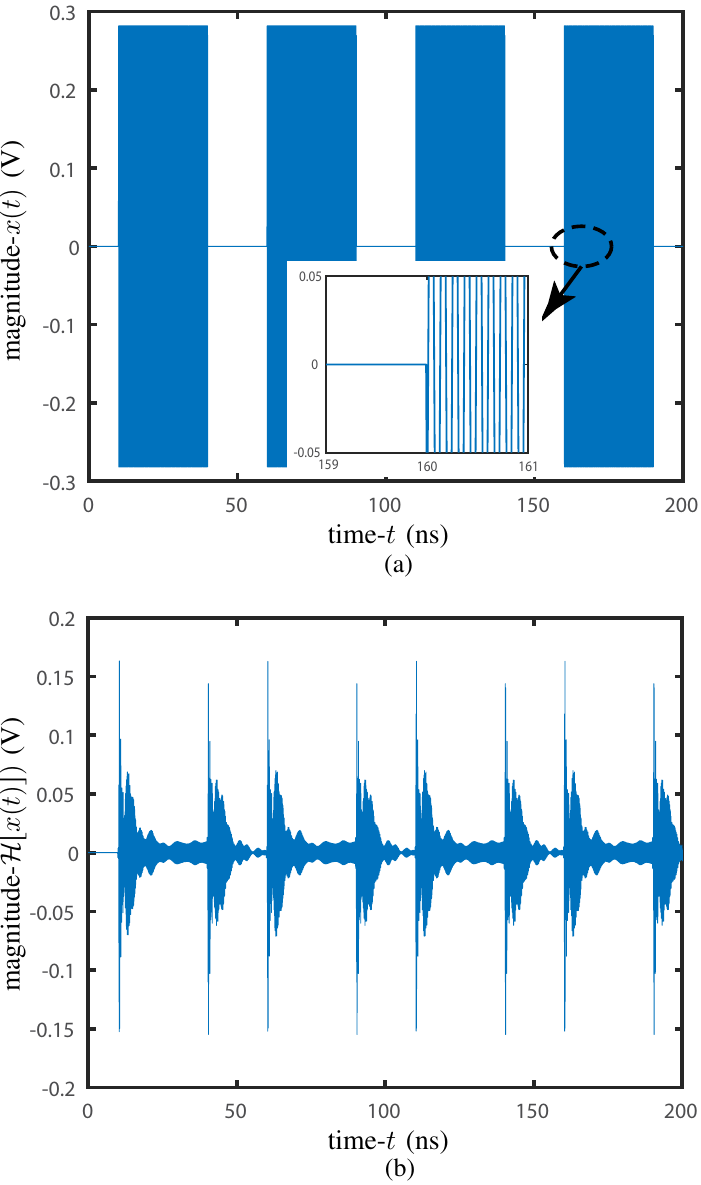}
       % \psfragfig[width=0.8\columnwidth, trim={0in 0in 0in 0in}]{Figs/Edge_Det}{
%        \psfrag{A}[c][c][0.8]{time -- $t$ (ns)}
%        \psfrag{B}[c][c][0.8]{magnitude -- $x(t)$ (V)}
%        \psfrag{C}[c][c][0.8]{magnitude -- $\mathcal{H}[x(t)])$ (V)}
%        \psfrag{a}[l][l][0.75]{(a)}
%        \psfrag{b}[l][l][0.75]{(b)}
%        }
        \caption{Edge detection verification for the RAP Hilbert transformer in \figref{FIG:HT_Exp}. (a)~Input rectangular pulse train modulated at $f_0=10$~GHz. (b)~Output signal with edge detection.}
   \label{FIG:Edge_Det}
\end{figure}

Figure~\ref{FIG:Edge_Det}(b) shows the response of the RAP Hilbert transformer in \figref{FIG:HT_Exp} to the modulated periodic rectangular pulse train in \figref{FIG:Edge_Det}(a). Since, in contrast to the case of \figref{FIG:ExpSquare}, the pulse is modulated, the detected edges have no specific polarity, but the experiment perfectly shows the edge detection operation of the Hilbert transformer.

\subsection{Peak Clipping}

The Hilbert-transformer is also a \emph{peak clipper}, and can thus be used as a peak-to-average ratio reducer in radar and switched power amplifiers, where the peak information can be recovered by a second Hilbert transformation, according to~\eqref{EQ:recov}. The clipping effect naturally exists in \figref{FIG:Edge_Det_Ideal}, where the center of the rectangular pulse is mapped onto zero, but the effect is not very significant there due to the flatness of the pulse.

\begin{figure}[h!t]
    \centering
    \includegraphics[width=0.78\columnwidth]{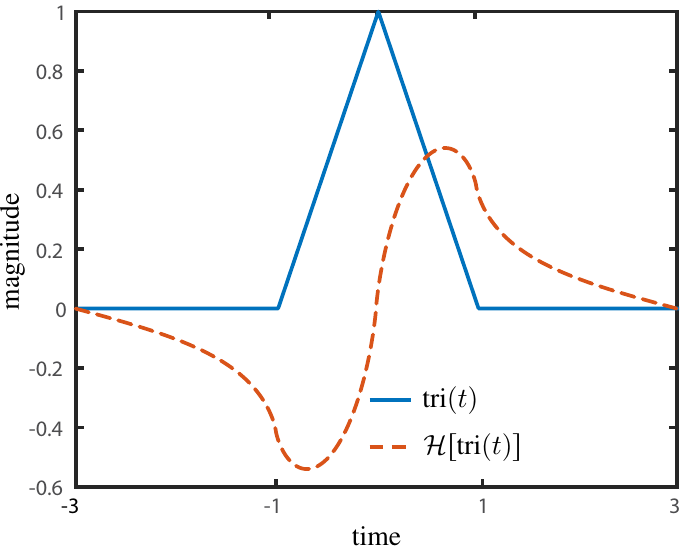}
        %\psfragfig[width=0.78\columnwidth, trim={0in 0in 0in 0in}]{Figs/Ideal_Hilbert_Triangle}{
%        \psfrag{a}[c][c][0.8]{time}
%        \psfrag{b}[c][c][0.8]{magnitude}
%        \psfrag{A}[l][l][0.8]{$\text{tri}(t)$}
%        \psfrag{B}[l][l][0.8]{$\mathcal{H}[\text{tri}(t)]$}
%        }
        \caption{Hilbert transform of a triangular pulse [Eqs.~\eqref{EQ:Tri} and~\eqref{EQ:Resp_of_Tri}].}
   \label{FIG:Peak_Det_Ideal}
\end{figure}

To better appreciate it, let us consider the triangular input signal
\begin{equation}\label{EQ:Tri}
x(t)=\text{tri}(t)=
\begin{cases}
  t+1,& \text{if}~ -1 \leq t < 0,\\
  -t+1,&  \text{if}~ 0\leq t \leq 1,\\
  0,& \text{otherwise}.
\end{cases}
\end{equation}
Computing the mathematical Hilbert transform of this function yields
\begin{equation}\label{EQ:Resp_of_Tri}
\begin{aligned}
  y(t)=&\frac{1}{\pi}\mathcal{P.V.}\int_{-\infty}^{+\infty}\text{tri}(\tau)\frac{1}{t-\tau}d\tau\\
      =&\frac{1}{\pi}\left(\mathcal{P.V.}\int_{-1}^{0}\frac{\tau+1}{t-\tau}d\tau+ \mathcal{P.V.}\int_{0}^{1}\frac{-\tau+1}{t-\tau}d\tau\right)\\
      =&\frac{1}{\pi}\left(\mathcal{P.V.}\int_{-1}^{0}\frac{t+1}{t-\tau}d\tau-\int_{-1}^{0}1d\tau+\mathcal{P.V.}\int_{0}^{1}\frac{1-t}{t-\tau}d\tau+\int_0^1 1d\tau\right)\\
      =&\frac{1}{\pi}\left(\mathcal{P.V.}\int_{t+1}^{t}\frac{t+1}{u}d(-u)+\mathcal{P.V.}\int_{t}^{t-1}\frac{1-t}{u}d(-u)\right)\\
      =&-\frac{1}{\pi}\left((t+1)\ln\left|\frac{t}{t+1}\right|+(1-t)\ln\left|\frac{t-1}{t}\right|\right)\\
      =&-\frac{1}{\pi}\left(\ln\left|\frac{t-1}{t+1}\right|+\frac{t}{1}\ln\left|\frac{t^2}{t^2-1}\right|\right)
      =\mathcal{H}[\Pi(t)],
\end{aligned}
\end{equation}
which maps $x(0)$ into $y(0)$, and thus indeed suppresses the peak value, as shown in \figref{FIG:Peak_Det_Ideal}
\vspace{1cm}

\begin{figure}[h!t]
    \centering
    \includegraphics[width=1\columnwidth]{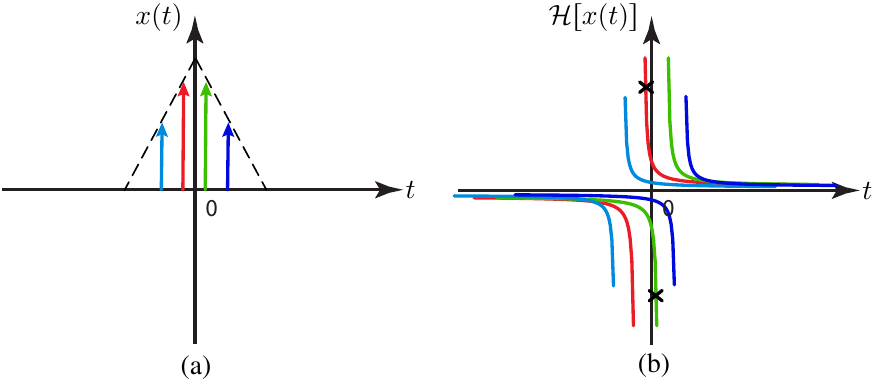}
        %\psfragfig[width=1\columnwidth, trim={0in 0in 0in 0in}]{Figs/ExpTriangle}{
%        \psfrag{a}[c][c][0.8]{(a)}
%        \psfrag{b}[c][c][0.8]{(b)}
%        \psfrag{w}[r][r][0.8]{$x(t)$}
%        \psfrag{x}[r][r][0.8]{$\mathcal{H}[x(t)]$}
%        \psfrag{1}[l][l][0.8]{$t$}
%        }
        \caption{Intuitively explanation of the triangular pulse peak suppression in \figref{FIG:Peak_Det_Ideal}. (a)~Input signal triangular pulse $x(t)=\text{tri}(t)$ expressed as a (continuous) superposition of triangularly weighted Dirac functions. (b)~Formation of the Hilbert transform of (a) using the fact that $y(t)=h(t)$ [Eq.~\eqref{EQ:ht}] for $x(t)=\delta(t)$.}
   \label{FIG:ExpTriangle}
\end{figure}

Figure~\ref{FIG:ExpTriangle} intuitively explains the peak suppression produced by the Hilbert transformer, following the same logics as for edge detection.

\begin{figure}[h!t]
    \centering
    \includegraphics[width=0.8\columnwidth]{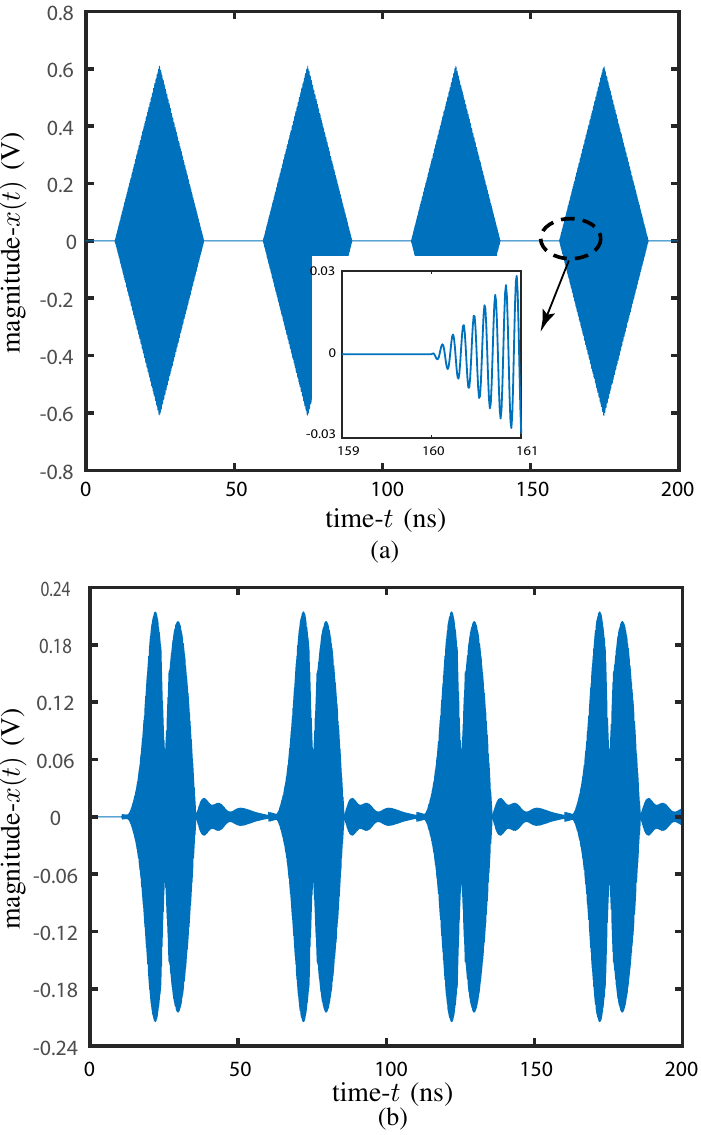}
       % \psfragfig[width=0.8\columnwidth, trim={0in 0in 0in 0in}]{Figs/peak_detection}{
%        \psfrag{B}[c][c][0.8]{time-$t$ (ns)}
%        \psfrag{A}[c][c][0.8]{magnitude-$x(t)$ (V)}
%        \psfrag{C}[c][c][0.8]{magnitude-$\mathcal{H}[x(t)])$ (V)}
%        \psfrag{a}[c][c][0.75]{(a)}
%        \psfrag{b}[c][c][0.75]{(b)}
      %  }
        \caption{Peak suppression verification for the RAP Hilbert transformer in \figref{FIG:HT_Exp}. (a)~Input triangular pulse train modulated at $f_0=10$~GHz. (b)~Output signal with peak suppression.}
   \label{FIG:Peak_Det}
\end{figure}

Finally, Figure~\ref{FIG:Peak_Det}(b) shows the response of the RAP Hilbert transformer in \figref{FIG:HT_Exp} to the modulated periodic triangular pulse train in \figref{FIG:Peak_Det}(a). Again, the response exhibits no specific polarity due to its modulated nature, but the experiment perfectly shows the peak suppression operation of the Hilbert transformer.

\subsection{Single-Sideband (SSB) Modulation}
\begin{figure}[h!t]
    \centering
    \includegraphics[width=1\columnwidth]{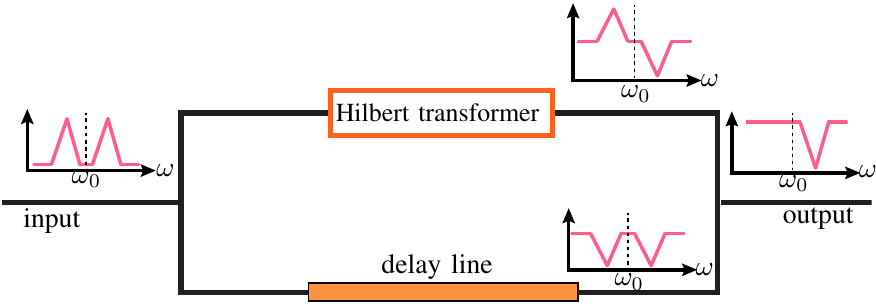}
        %\psfragfig*[width=1\columnwidth, trim={0in 0in 0in 0in}]{Figs/SSM}{
%        \psfrag{A}[c][c][0.75]{Hilbert transformer}
%        \psfrag{B}[c][c][0.8]{delay line}
%        \psfrag{a}[l][l][0.8]{input}
%        \psfrag{b}[r][r][0.8]{output}
%        \psfrag{t}[c][c][0.8]{$\omega$}
%         \psfrag{w}[c][c][0.8]{$\omega_0$}
%        }
        \caption{Schematic of a SSB modulation based on a RAP Hilbert transformer.}
   \label{FIG:SSM}
\end{figure}

The Hilbert transformer may be combined with a delay line to provide a single side-band (SSB) modulator, as shown in \figref{FIG:SSM}. In this application, the transformer center frequency is set at the middle of the two initial sidebands. It reverses the sign of one of the two sidebands and combines the result with a delayed copy of the input signal, as shown in the figure, which results in the suppression of the undesired sideband. The effect is identical to that of a low-pass (lower SSB) or high-pass (upper SSB) filter, but this implementation allows much easier SSB switching by simply tuning the length of the transmission line using a U-shaped detour loaded by a PIN diode.

\begin{figure}[h!t]
    \centering
    \includegraphics[width=0.8\columnwidth]{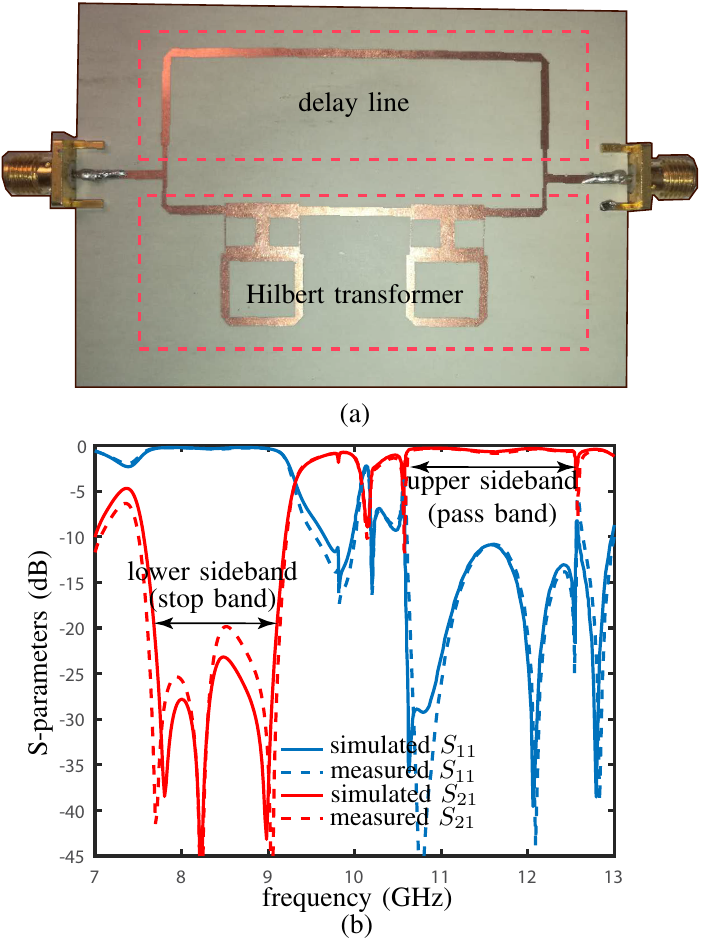}
        %\psfragfig*[width=0.8\columnwidth, trim={0in 0in 0in 0in}]{Figs/SSM_Exp}{
%        \psfrag{A}[c][c][0.8]{frequency (GHz)}
%        \psfrag{B}[c][c][0.8]{S-parameters (dB)}
%        \psfrag{a}[l][l][0.75]{simulated $S_{11}$}
%        \psfrag{b}[l][l][0.75]{measured $S_{11}$}
%        \psfrag{c}[l][l][0.75]{simulated $S_{21}$}
%        \psfrag{d}[l][l][0.75]{measured $S_{21}$}
%        \psfrag{e}[c][c][0.8]{Hilbert transformer}
%        \psfrag{f}[c][c][0.8]{delay line}
%        \psfrag{g}[c][c][0.8]{(a)}
%        \psfrag{h}[c][c][0.8]{(b)}
%        \psfrag{C}[c][c][0.8]{\shortstack{lower sideband\\ (stop band)}}
%        \psfrag{D}[c][c][0.8]{\shortstack{upper sideband \\ (pass band)}}
%        }
        \caption{Experiment demonstration of a SSB modulator based on a RAP Hilbert transformer. (a)~Fabricated prototype, for upper SSB operation. (b)~Simulated and measured high-pass filtering S-parameters.}
   \label{FIG:SSB_EXP}
\end{figure}

The response of the Hilbert transformer based SSB modulator is experimentally demonstrated in~\figref{FIG:SSB_EXP}. The sideband switching operation is trivial and is hence not presented here.

\section{Conclusion}
A microwave Hilbert transformer, based on the combination of a branch-line coupler and a loop resonator, has been introduced, characterized, physically explained, and applied to edge detection, peak suppression and single side-band modulation.

The device represents a new component for Real-time Analog Processing (RAP), and is hence expected to help the development of this technology in the forthcoming years.

\bibliographystyle{IEEEtran}
\bibliography{IEEEabrv,Xiaoyi_Reference}
\end{document}